\documentclass[preprint2]{aastex}

\shorttitle{Neutrino Cooled Accretion Disks and Short GRBs}

\shortauthors{Lee, Ramirez-Ruiz \& Page}

\begin{document}


\title{Opaque or transparent? A link between neutrino optical depths
and the characteristic duration of short Gamma Ray Bursts}


\author{William H. Lee}
\affil{Instituto de Astronom\'{\i}a, 
Universidad Nacional Aut\'{o}noma de M\'{e}xico, \\ 
Apdo. Postal 70-264, Cd. Universitaria, 
M\'{e}xico D.F. 04510}

\author{Enrico Ramirez--Ruiz\footnote{Chandra 
Fellow}} 
\affil{School of Natural Sciences, Institute for Advanced Study, 
Princeton, NJ, 08540 }

\and

\author{Dany Page}
\affil{Instituto de Astronom\'{\i}a, 
Universidad Nacional Aut\'{o}noma de M\'{e}xico, \\ 
Apdo. Postal 70-264, Cd. Universitaria, 
M\'{e}xico D.F. 04510}



\begin{abstract}
Cosmological gamma ray bursts (GRBs) are thought to occur from violent
hypercritical accretion onto stellar mass black holes, either
following core collapse in massive stars or compact binary
mergers. This dichotomy may be reflected in the two classes of bursts
having different durations.  Dynamical calculations of the evolution
of these systems are essential if one is to establish characteristic,
relevant timescales. We show here for the first time the result of
dynamical simulations, lasting approximately one second, of
post--merger accretion disks around black holes, using a realistic
equation of state and considering neutrino emission processes. We find
that the inclusion of neutrino optical depth effects produces
important qualitative temporal and spatial transitions in the
evolution and structure of the disk, which may directly reflect upon
the duration and variability of short GRBs.
\end{abstract}




\section{Introduction}\label{intro}

The coalescence of two compact objects due to the emission of
gravitational radiation is a possible origin of cosmological gamma ray
bursts of the short variety \citep[see e.g.,][for a review]{m02},
lasting a few tenths of a second
\citep{ls76,bp86,elps89,npp92}. Previous multidimensional studies have
shown that the outcome of such a merger is in all likelihood a dense
torus surrounding a supramassive neutron star, likely to collapse to a
black hole on a short timescale, or a black hole, if one was already
present in the system \citep[e.g][]{rs94,kl98,rj99,rrd03}. This class
of systems has been studied observationally since the 1970s, when the
Hulse--Taylor system was discovered \citep{ht75}, and the recent
detection of PSR J0737-3039 \citep{burgay03}, the tightest binary yet
in this class, has only heightened interest in these events. The final
instants before the actual merger will produce a powerful burst of
gravitational waves, and the merger waveform itself is expected to
reveal details about the equation of state of matter at high densities
\citep[see e.g.,][]{thorne95}.

The merger process is intrinsically a three--dimensional event, with
no particular symmetry which can be exploited to reduce the complexity
of the problem, and must be addressed numerically. No such calculation
is currently evolved for more than a few tens of milliseconds, because
of computational limitations. Information about the state of the
system after the violent, dynamical merger is essentially over can be
obtained, but the final state of the system, on a timescale comparable
to the duration of a short GRB, cannot be addressed.

The steady state structure and composition of post--merger accretion
disks has been the focus of several recent studies
\citep{pwf99,npk01,km02,dpn02,pwh03,b03}, which have included an increasing
amount of microphysical detail. However, the dynamics of the problem
have never been followed for an extended amount of time, although
progress is being made in this direction \citep{srj04}.

In a previous paper \citep{lrr02}, we studied the dynamical evolution
of the accretion structures formed after a black hole--neutron star
merger under a set of simplifying assumptions. Initial conditions were
taken directly from the late stages of the coalescence, modeled in
three dimensions using a simple equation of state \citep{l01}. The
disk was then mapped onto two dimensions assuming azimuthal symmetry,
and its evolution was followed for 0.2 seconds, comparable to the
duration of a short GRB. The equation of state was simple (ideal gas
with adiabatic index $\gamma=4/3$), and we assumed that all the energy
dissipated by viscosity (modeled with the full expression of the
stress tensor, using an $\alpha$--prescription for the magnitude of
the viscosity) was radiated away.

In this {\em Letter} we present for the first time dynamical
calculations of post--merger accretion disks in two dimensions,
$(r,z)$, that: (1) use a realistic and self--consistent equation of
state for the fluid in the disk; (2) consider the cooling of the disk
through neutrino emission, using appropriate rates; (3) account for
the effects of neutrino opacities and the corresponding optical depths
through a simplified treatment and (4) last for a minimum of 0.4
seconds, comparable to the duration of a short GRB.

\section{Physics}\label{physics}

After the initial merger phase of the binary is completed, a few
tenths of a solar mass can be left in orbit around the central black
hole (which has a mass $M_{\rm BH}\approx$~3--5~$M_{\odot}$), in a
small disk a few hundred kilometers across. The densities are high,
with $10^{9}\leq\rho \mbox{(g~cm$^{-3}$)}\leq 10^{12}$, and the
corresponding temperatures are $10^{10}\leq T \mbox{(K)}\leq10^{11}$.
Under these conditions, nuclei are entirely photodisintegrated, and
the fluid consists of free electrons, protons, nucleons, plus a small
number of positrons. The nucleons make up a non--relativistic ideal
gas, while electrons are degenerate and highly relativistic. Photons
are completely trapped due to the extremely high optical depth, and
are hence advected with the flow. The composition of the fluid is
determined self--consistently assuming: (1) charge neutrality and (2)
complete $\beta$--equilibrium \citep{km02,lrrp04}. This allows us to
determine the electron fraction $Y_{e}$ and the electron degeneracy
parameter (through the chemical potential) locally, and to follow it
throughout the dynamical evolution. This is important since the
composition directly affects the neutrino emission rates (see below).
The equation of state is then given by
\begin{equation}
P=\frac{aT^{4}}{3}+\frac{\rho k T}{m_{u}}+K\left(
\frac{\rho}{\mu_{e}}\right)^{4/3}+P_{\nu},
\end{equation}
where $K$ is the constant corresponding to a fully degenerate,
relativistic electron gas, $\mu_{e}=1/Y_{e}$ is the number of baryons
per electron, and $P_{\nu}$ is the pressure due to the presence of
neutrinos in the optically thick regions of the disk (see
below). Strictly speaking, the electrons are not fully degenerate,
since $\eta_{e}/kT$ is in the range 2--4 ($\eta_{e}$ is the electron
chemical potential). However, this represents a small error in the
equation of state, since gas pressure dominates the overall balance,
contributing about 90\% of the total pressure, followed by electron
degeneracy pressure at around 5\%.

Neutrinos are emitted abundantly in the disk, due to the high
temperatures and densities. We take into account all of the following
processes: (1) nucleon--nucleon bremmsstrahlung; (2) $e^{\pm}$
pair annihilation; (3) electron and positron captures onto free
nucleons and (4) plasmon decay. In practice, the global cooling rate
is completely dominated by $e^{\pm}$ captures, and is hence well
approximated by \citep{km02}:
\begin{equation}
\dot{q}_{cap}=1.1\times 10^{31} \, \left[
\frac{\eta_{e}}{kT}\right]^{9} \, T_{11}^{9} \,
\mbox{erg~s$^{-1}$~cm$^{-3}$}\label{qcap2}
\end{equation}
in the regime where electrons are degenerate. This expression has been
used directly to compute the cooling within the disk. The error in
doing so, as opposed to a full integration over the phase space, is
limited to about a factor of 2 (in the inner regions, where the
cooling is most intense), as compared with rates extrapolated from the
work of \citet{lmp01}.

The emitted neutrinos are not entirely free to leave the system, since
scattering off free nucleons is important. The optical depth for
neutrinos can be estimated as $H/l_{\nu}$, where $H$ is a typical disk
scale height, and $l_{\nu}$ is the mean free path between
scatterings. The cross section depends on the mean neutrino energy as
$\sigma_{n}\propto E_{\nu}^{2}$ \citep{ts75}, and $E_{\nu}\approx 43
(Y_{e}\rho_{12})^{1/3}$~MeV ($\rho_{12}=\rho/10^{12}$~g~cm$^{-3}$),
using the fact that neutrinos are being produced by reactions
involving degenerate electrons \citep[see e.g.,][]{st83}. This amounts
to a surface of last scattering for neutrinos, or
``neutrino--surface'' at $\rho \simeq 10^{11}$g~cm$^{-3}$. Our simple
treatment of this fact consists of suppressing the local cooling rate
by a factor $\exp(-\tau_{\nu})$, and adding a pressure term
$P_{\nu}\propto aT^{4}[1-\exp(-\tau_{\nu})]$ to the equation of
state. This simple alteration is crucial in determining the energy
output of the disk. We note that no modification in the composition
was computed owing to this fact. In reality, the opaqueness of the
material will lead to an increase in the electron fraction $Y_{e}$, by
up to a factor of 2 \citep{b03}. The optical depth will thus rise by a
factor $2^{2/3}\simeq 1.6$, making the optically thick region in the
disk larger and the overall luminosity slightly lower. Any energy
release in neutrinos will thus be spread over a longer period, making
our estimates concerning time scales fall on the conservative side
(see below, Section~\ref{transitions}).

The dynamical evolution is followed for at least 0.4 seconds in the
central Newtonian potential produced by the black hole, $\Phi=-GM_{\rm
BH}/R$, using a two dimensional smooth particle hydrodynamics (SPH)
code in azimuthal symmetry \citep{monaghan92}. Accretion is modeled by
placing an absorbing boundary at the Schwarszchild radius
$r_{G}=2GM_{\rm BH}/c^{2}$, and the mass of the black hole is updated
continuously. There is no external agent feeding the disk with matter,
and no boundary conditions need to be specified (other than for
accretion). The equations of motion and the energy equation contain
all the terms for a physical viscosity derived from the stress tensor
\citep{flebbe94}, and we use an $\alpha$--prescription for the
coefficient of viscosity, with $ 10^{-3}<\alpha < 10^{-1}$. The
dissipated energy is injected back into the disk as thermal energy,
and may leave the system depending on the local cooling rate (see
above).

\section{Temporal and spatial transitions}\label{transitions}

\begin{figure}
\plotone{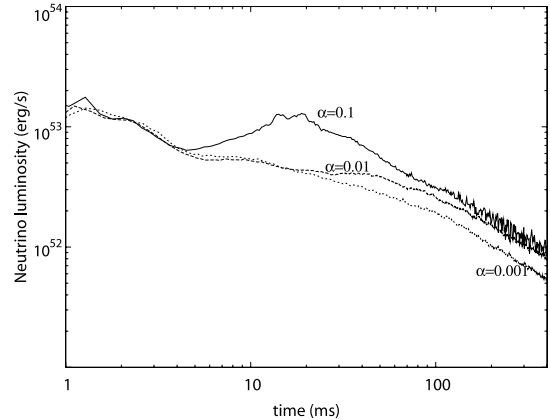}
\caption{Neutrino luminosity as a function of time for three runs
using identical initial conditions and different values of the
viscosity parameter: $\alpha=10^{-1}$, $\alpha=10^{-2}$ and
$\alpha=10^{-3}$.
\label{fig1}}
\end{figure}

If the disk is assumed to be optically thin everywhere, and the
effects discussed above are ignored, the neutrino luminosity is a
smooth, monotonically decreasing function of time, similar to what we
obtained in previous work (we have confirmed this in trial runs with
the present input physics). Accounting for a variable optical depth
leads to behavior that is markedly different, as can be seen in
Figure~\ref{fig1}, where $L_{\nu}(t)$ is shown for three values of the
viscosity parameter $\alpha$. For a high viscosity, $\alpha=0.1$, the
accretion timescale is so short (40~ms), that most of the material is
accreted onto the black hole and has no time to radiate away its
internal energy reservoir of $\approx 10^{52}$~erg. The disk becomes
entirely optically thin at around 30~ms, and this explains the
rebrightening that is observed. The luminosity subsequently exhibits a
power law decay, as in the trial runs with no optical depth effects,
with $L_{\nu}\propto t^{-0.9}$.

For lower viscosities ($\alpha=10^{-2}, 10^{-3}$), the behavior is
qualitatively different at early times, and is a consequence of the
longer accretion timescales: 0.4~s and 1~s respectively (we note here
that all our estimates of $t_{\rm acc}$ are based on the actual
accretion rate, i.e. $t_{\rm acc}\approx M_{\rm d}/\dot{M}_{\rm
BH}$). Quite simply the material in the disk is allowed to remain in
the vicinity of the black hole without being accreted, at high
temperatures and densities (since the transport of angular momentum is
so inefficient), and radiate away essentially all of its internal
energy through the emission of neutrinos. Once this occurs, on a
cooling timescale given by $t_{\rm cool}\approx E_{\rm int}/L_{\nu}$,
a break occurs (at $t\simeq 80, 100$~ms for $\alpha=10^{-2}, 10^{-3}$
respectively) and the luminosity drops again as $t^{-0.9}$. At this
stage the energy output comes directly from viscous dissipation within
the disk. The break time is determined by how much energy the disk
holds initially, $E_{\rm int}$, and by the rate at which it is lost,
$L_{\nu}(t)$, which is itself fixed by the densities and temperatures
found within it. The limitation on the cooling rate imposed by high
optical depths is essential, and allows the energy loss to be spread
over an extended period of time. For the three calculations shown
here, the total energy release in neutrinos, up to 0.4~s, is
$(E_{\nu}/10^{52}$~erg)$=1.2, 1.05, 0.8$, for $\alpha=10^{-1},
10^{-2}, 10^{-3}$, respectively. Assuming that this energy is
converted to $e^{\pm}$ with a one per cent efficiency in the regions
along the rotation axis through $\nu \overline{\nu}$ annihilation,
approximately 10$^{50}$~erg could be released and made available for
the production of a relativistic fireball. A detailed calculation of
this is left as future work.

\begin{figure} 
\plotone{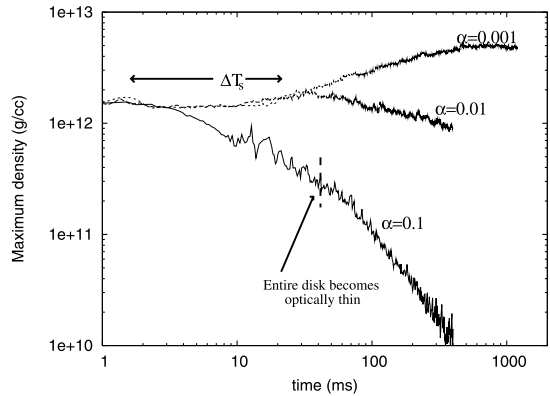} 
\caption{Maximum density as a function of time for the same runs as
shown in Figure~\ref{fig1}. The interval marked $\Delta T_{s}$
corresponds to the sound crossing time across the optically thick
region of the disk. The subsequent rise is due to the vertical
compression of the disk as it cools. Accretion eventually drains the
disk of matter, producing the characteristic breaks at $t=0.05, 0.8$~s
for $\alpha=10^{-2}, 10^{-3}$.
\label{fig2}} 
\end{figure}

The other important mechanism by which a GRB might be powered is MHD
energy extraction, either through the Blandford--Znajek mechanism
\citep{bz77}, or through a magnetically dominated jet. Since our code
does not explicitly account for the presence of magnetic fields, we
make a simple, rough estimate of the magnetic field strength by
assuming that its energy density, $B^{2}/8\pi$, is in equiparition
with the internal energy density, $\rho c_{s}^{2}$ (this is clearly an
upper limit).  
Within the disk, the field will be most intense where the density is
highest. Figure~\ref{fig2} shows the maximum density in the disk as a
function of time, for the same calculations as shown in
Figure~\ref{fig1}. The inferred magnetic field is $B \simeq
10^{15}-10^{16}$~G. It is again evident that the high--viscosity case
is markedly different from the other two, showing a significant prompt
decrease in $\rho_{\rm max}$. The transition to complete transparency
to neutrinos occurs at around 40~ms. The balance between cooling,
which produces vertical compression (and hence a rise in central
density), and accretion, which slowly reduces the mass of the disk,
produces a break in the curves for $\alpha=0.01, 0.001$ at $t \approx
0.05, 0.8$~s respectively.  The total estimated energy release through
the Blandford--Znajek mechanism is $(E_{\rm BZ}/10^{51}$~erg)$=.34,
1.3, 6.5$, for $\alpha=10^{-1}$ (up to 0.4~s), $\alpha=10^{-2}$ (up to
0.4~s), $\alpha=10^{-3}$ (up to 1.2~s), respectively.

\begin{figure}[h!]
\plotone{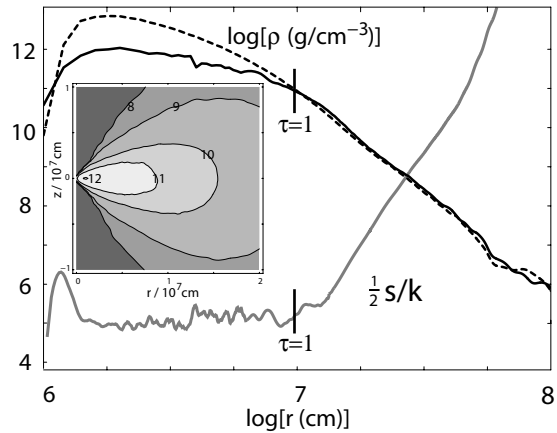}
\caption{Density profile and entropy per baryon along the equatorial
plane, $z=0$, 0.2~s after the start of the calculation with
$\alpha=10^{-2}$ (solid line). The dashed line indicates the density
profile for the corresponding trial run that did not include optical
depth effects. The vertical mark on each curve indicates the position
where $\tau_{\nu}=1$. {\it Inset Panel:} Logarithmic density contours
(equally spaced every dex). The neutrino--surface corresponds roughly
to $\log \rho=10^{11}$~g~cm$^{-3}$. Under these circumstances, the gas
cannot efficiently cool and mantains a pressure scale height
comparable with its radius. This, however, cannot happen too close to
the symmetry axis where the centrifugal force cannot oppose the
vertical pull of gravity and a funnel is formed.
\label{fig3}}
\end{figure}

In addition to the temporal variations in total luminosity, there is a
spatial transition within the disk, when the optical depth is around
unity. For the calculations with $\alpha \leq 10^{-2}$, there is
always (up to the end of the run) an opaque region in the center of
the disk. The density $\rho$ and entropy per baryon $s/k$ along the
equatorial plane ($z=0$), as well as density contours in the $r-z$
plane, are plotted in Figure~\ref{fig3}, 0.2~s after the beginning of
the calculation with $\alpha=10^{-2}$. We also show for comparison the
equatorial density profile of the disk in a calculation that did not
take into account the effects of optical depths (dashed line in
Figure~\ref{fig3}). There is a clear break at $r\approx 9 \times
10^{6}$~cm, where $\tau_{\nu}=1$. As the optical depth increases and
the cooling is suppressed, the pressure rises, inhibiting the thinning
of the disk and the increase in density in the inner regions. This
break is essentially a stationary feature, due to the relatively long
accretion time scales, as long as the optically thick region is
present. We believe this transition is significant, as geometrically
thick accretion flows around rotating black holes, with $H\simeq R$,
are believed to be more favorable for the production of energetic
outflows than thin flows, where $H \ll R$ \citep{lop99,meier01}.

\section{Discussion}\label{disc}

The evolution of accretion disks resulting from dynamical three
dimensional binary coalescence calculations, where a neutron star is
tidally disrupted before being swallowed by its black hole companion,
is studied numerically. Angular momentum transport and the associated
energy dissipation are modeled using an $\alpha$ prescription. By
assuming azimuthal symmetry we are able to follow the time dependence
of the disk structure for a fraction of a second, a time comparable
with the duration of short GRBs. During this time, the rate of mass
supply to the central black hole is of the order of a fraction of a
solar mass per second; i.e. much greater than the Eddington
rate. Although the gas photon opacities are large, the disk becomes
sufficiently dense and hot to cool via neutrino emission. There is in
principle no difficulty in dissipating the disk internal energy, but
the problem is in allowing these neutrinos to escape from the Thomson
thick inflowing gas.

At sufficiently low accretion rates, $\alpha \lesssim 0.01$, we find
that the energy released by viscous dissipation is almost completely
radiated away on a timescale given by $t_{\rm cool} \approx E_{\rm
int}/L_{\nu} \sim 0.1$~s. In contrast, for a higher mass supply,
$\alpha \gtrsim 0.1$, energy advection remains important until the
entire disk becomes optically thin, at $t[\tau_\nu=1]\sim 30$ ms. The
restriction on the cooling rate imposed by high optical depths is key
because it allows the energy loss to be spread over an extended period
of time (i.e.  $t_{\rm cool}$ or $t[\tau_\nu=1]$) during which the
neutrino luminosity stays roughly constant.  In principle, neutrinos
could give rise to a relativistic pair dominated wind if they
converted into $e^\pm$ pairs in a region of low baryon density. This
gives a characteristic timescale for energy extraction, and may be
essential for determining the duration of neutrino-driven, short GRBs.

An alternative way to tap the torus energy is via magnetic fields
threading the disk: the energy liberated by accretion is converted
efficiently into magnetic form and emitted as a magnetically dominated
outflow. The launching of a jet probably requires the existence of a
poloidal magnetic field of magnitude $B_p$ over a scale of radius $R$,
where $B_p$ is smaller than the field strength $B_{\rm disk}$
associated with the dynamo-driven magnetic disk viscosity
\citep{lop99}. This results directly from estimating the accretion and
jet luminosities as $L_{\rm acc}\sim\dot{M}_{\rm acc}v_{\phi}^{2}$,
and $L_{\rm jet}\sim B_{\rm p}^{2}R^{2}v_{\phi}$ respectively, which
gives $(B_{\rm p}/B_{\rm disk})^{2}\sim (L_{\rm jet}/L_{\rm
acc})(H/R)$. Assuming the scaling $B_{\rm p}\sim (H/R)B_{\rm disk}$,
derived by \citet{tp96}, one obtains $L_{\rm jet}/L_{\rm acc}\sim
H/R$. In order to achieve $L_{\rm jet} \sim L_{\rm acc}$,
\citet{lpk03} argue that a number $R/H$ of neighboring annuli in the
disk need to provide a locally generated net poloidal field in the
same direction. They notice that since the local dynamos vary on
timescales of $t_{\rm dyn}$, the timescale for establishing a change
in the magnetic field in the disk should be of the order $t_{\rm jet}
\sim t_{\rm dyn}2^{R/H}$. In our case $R/H\sim 3$, which gives $t_{\rm
jet}\sim 10-100$ ms. This corresponds to the typical variability
timescale of short duration GRBs \citep{rf00,np02}.

Two further physical effects, noted previously in a more general
context, are relevant for the present study. First, the mechanism by
which a jet is launched is not scale--free \citep{pringle93}, and thus
the poloidal field must be established in the inner regions of the
disk. Second, in these same regions, the inflow speed caused by
angular momentum loss to the outflow is increased, up to a factor
$R/H$ over the value expected from magnetic viscosity alone
\citep{lpk03}. Thus the poloidal flux generated by the dynamo can be
effectively trapped. This probably implies that once the dynamo
process generates a global poloidal field, it is able to maintain it
during the lifetime of the thick portion of the disk,
$t[\tau_\nu=1]$. For the typical parameters studied here, this
timescale could be of the order of seconds when the rate of mass
supply is low (i.e. $\alpha\sim 10^{-3}$). There is thus no problem in
principle in accounting for sporadic large-amplitude variability on
timescales as short as one millisecond, even in the most long-lived
short GRBs.

\acknowledgments

We gratefully acknowledge helpful discussions with M. Rees and W.
Klu\'{z}niak. We thank the referee, Jason Pruet, for his comments and
suggestions. Financial support for this work was provided in part by
CONACyT (36632E) and NASA through a Chandra Postdoctoral Fellowship
award PF3-40028 (ER-R).

\end{document}